\documentclass[superscriptaddress, letterpaper,twocolumn,prl,showpacs, showkeys,reprint]{revtex4}
\usepackage{amsmath, amsthm, amssymb}
\usepackage[pdftex]{graphicx}
\usepackage{dcolumn} 
\usepackage{bm} 

\makeatletter 
\gdef\@ptsize{2}
\let\@currsize\normalsize 
\makeatother 
\usepackage{setspace} 
\usepackage[usenames,dvipsnames]{color}

\begin{document}
\title{An efficient and reliable growth method for epitaxial complex oxide films by molecular beam epitaxy}

\author{T.W. Zhang}
\author{Z.W. Mao}
\author{Z.B. Gu}
\email{zbgu@nju.edu.cn}
\affiliation{National Laboratory of Solid State Microstructures,  College of Engineering and Applied Sciences, and Collaborative Innovation Center of Advanced Microstructures, Nanjing University, Nanjing 210093, China}
\author{Y.F. Nie}
\email{ynie@nju.edu.cn}
\affiliation{National Laboratory of Solid State Microstructures,  College of Engineering and Applied Sciences, and Collaborative Innovation Center of Advanced Microstructures, Nanjing University, Nanjing 210093, China}
\author{X.Q. Pan}
\affiliation{National Laboratory of Solid State Microstructures,  College of Engineering and Applied Sciences, and Collaborative Innovation Center of Advanced Microstructures, Nanjing University, Nanjing 210093, China}
\affiliation{Department of Chemical Engineering and Materials Science and Department of Physics and Astronomy, University of California, Irvine, 916 Engineering Tower, Irvine, CA 92697, USA}



\begin{abstract}

{\bf 
Transition metal oxide heterostructures and interfaces host a variety of exciting quantum phases and can be grown with atomic-scale precision by utilising the intensity oscillations of $in$ $situ$ reflection high-energy electron diffraction (RHEED). However, establishing a stable oscillation pattern in the growth calibration of complex oxides films is very challenging and time consuming. Here, we develop a substantially more efficient and reliable growth calibration method for complex oxide films using molecular beam epitaxy. 
 }
\end{abstract}

\maketitle


Transition metal oxide heterostructures and interfaces exhibit a wide variety of exotic correlated quantum phases and hold the promise of exciting electronic applications~\cite{Hwang:2012hz,Ohtomo2004, Chakhalian:2012hu,Reyren31082007}. High quality oxide heterostructures and interfaces can be fabricated with atomic-scale precision using advanced growth techniques with the application of reflection high-energy electron diffraction (RHEED). RHEED intensity oscillations reflect the film growth kinetics and the period corresponds to the growth of a repeat unit (e.g. an unit cell)~\cite{Haeni2000, brooks2009growth, nie2014atomically, lee2013exploiting, haeni2004room, wang2012cation}. To achieve most atomically sharp heterostructures and interfaces, a method that can precisely deposit one monolayer at a time would be preferred if the growth parameters can be calibrated precisely. SrTiO$_3$ is a prototype perovskite oxide that can be grown at a wide range of temperature and partial pressures of oxygen and on various templates including Silicon~\cite{Ohtomo2004, haeni2004room,li2014fabricating,warusawithana2009ferroelectric, Bhuiyan1, Bhuiyan2}. In the growth of SrTiO$_3$ films using a shuttered method by molecular beam epitaxy (MBE), SrO and TiO$_2$ monolayers are deposited alternatively and the film is grown in a layer-by-layer manner. Typically, the RHEED intensity increases as the growth of SrO layer and decreases as the growth of TiO$_2$ layer~\cite{Haeni2000, brooks2009growth} although opposite phenomena were also reported~\cite{sullivan2015complex}. During the calibration process, the precise shutter times of Sr and Ti sources for the deposition of full SrO and TiO$_2$ monolayers are obtained by stabilizing the RHEED intensity oscillation till it shows no clear amplitude change or overall intensity drift. Using these calibrated shutter times, high quality SrTiO$_3$ films and SrTiO$_3$ based superlattices and interfaces can thus be grown~\cite{Haeni2000, brooks2009growth, nie2014atomically, lee2013exploiting}. However, the RHEED oscillation patterns of complex oxides are very complicated and minor variations of the electron beam incident angle or shutter times can result in significant deviations~\cite{Haeni2000, sullivan2015complex, brooks2009growth, lei2016constructing}. Therefore, it is very challenging and time consuming to establish a stable RHEED oscillation pattern in calibrating the growth parameters of complex oxides, hindering the growth of more sophisticated oxide heterostructures and interfaces. 



In this letter, we show that the co-deposition method, a technique deposing atoms of all species simultaneously, is substantially more efficient and reliable than the conventional shuttered method in calibrating the growth parameters for complex oxide films. 


Epitaxial SrTiO$_3$ and Ruddlesden-Popper (RP) layered strontium titanate (Sr$_{n+1}$Ti$_n$O$_{3n+1}$) films were grown on (001) SrTiO$_3$ single-crystalline substrates at 750 $^{\circ}$C and 5$\times 10^{-7}$ Torr of a mixed oxidant (10\%O$_3$+90\%O$_2$)  using a DCA R450 MBE system. SrTiO$_3$ substrates were etched in buffered HF acid at room temperature and annealed in oxygen at 950$^{\circ}$C for 80 minutes to obtain a TiO$_2$-terminated step-and-terrace surface~\cite{kawasaki1994atomic}. A co-deposition method was used for calibrating the growth parameters and a shuttered method was employed in the final film growth. In the shuttered growth mode, Sr and Ti atoms were deposited one after the other in a computer-controlled sequence, while Sr and Ti atoms were deposited simultaneously in the co-deposition mode. After the calibration process, new substrates were loaded and kept rotating throughout the growth process in order to improve the uniformity of real samples. An electron beam of 15 keV energy was employed for the RHEED measurements. The film crystalline structure was examined by high-resolution x-ray diffraction (XRD) using a Bruker D8 Discover diffractometer.


\begin{figure}
\begin{center}
\includegraphics{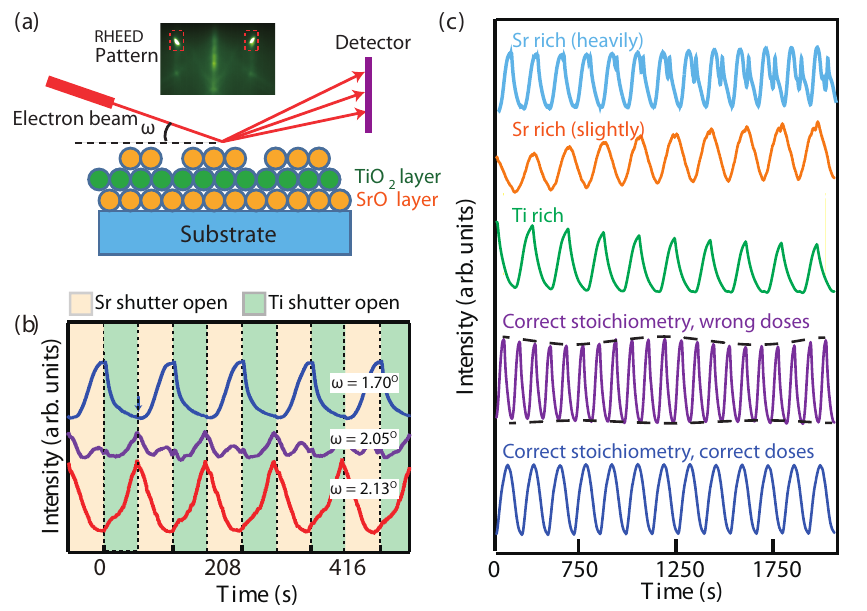}
\caption{(Color online) (a) Schematic of a typical RHEED system. The intensity variations of diffraction peaks enclosed by dashed lines reflect the kinetics of film growth.  (b) Sensitive dependance of shuttered RHEED oscillations on the incident angle of the electron beam in the growth of SrTiO$_3$ films on (001) SrTiO$_3$ substrate.  (c) Complicated shuttered RHEED oscillation patterns as a function of stoichiometry and monolayer doses.
\label{fig:TypicalRHEED}}
\end{center}
\end{figure}

Figure 1 shows the complicated shuttered RHEED oscillations of (1 1) diffraction peak in the MBE growth of SrTiO$_3$ films, showing clear and strong dependance on the electron beam incident angle (Fig. 1b) and the accuracy of shutter times (Fig. 1c). Similar incident angle dependance has been reported in the RHEED oscillation by monitoring the (0 0) streak but its origin is still not fully understood~\cite{Haeni2000}. Consistent with literatures~\cite{Haeni2000, brooks2009growth, lei2016constructing}, the overall intensity increases (decreases) or even forms spitting "double peaks" pattern around its maximum (minimum) intensity in Sr-rich (Ti-rich) condition. A beating pattern can be formed if the stoichiometry is correct but the monolayer doses are off. Eliminating this beating pattern is necessary for precise film growth but it is very time consuming due to the long beating period as the stoichiometry approaches its ideal value. Stable RHEED oscillation pattern can only be achieved when Sr and Ti shutter times are both accurate. A misleading false "stable" RHEED pattern can be formed when the beating becomes very weak as the film surface becomes rough. All together, these complicated RHEED oscillation patterns and the sensitive incident angle dependence make the shuttered method unfavourable in calibrating the growth parameters for complex oxide heterostructures and interfaces.

In contrast to the shuttered method, the RHEED oscillation pattern in a co-deposition process is simple and clear, which is most likely due to the less abrupt flux discontinuities. In Fig. 2a, we show a typical calibration process by co-deposition method where both Sr and Ti atoms are deposited on the substrate simultaneously. Only a monotonic drift of the overall intensity of the oscillations with no "double peaks", beating or other complicated patterns are observed. No clear dependence of the incident angle is observed neither. Similar to the shuttered method, the overall oscillation intensity increases in Sr-rich condition and decreases in Sr-poor condition. By adjusting Sr (or Ti) source temperature (flux rate), the monotonic intensity drift can be eliminated. Within a few tens of oscillation periods, a stable RHEED oscillation pattern was established and the period corresponds to the correct shutter times for the growth of a full monolayer of SrO and TiO$_2$. Therefore, this co-deposition calibration process is significantly more reliable and efficient than the shuttered method by reducing the calibration time by at least an order of magnitude. 


\begin{figure}
\begin{center}
\includegraphics {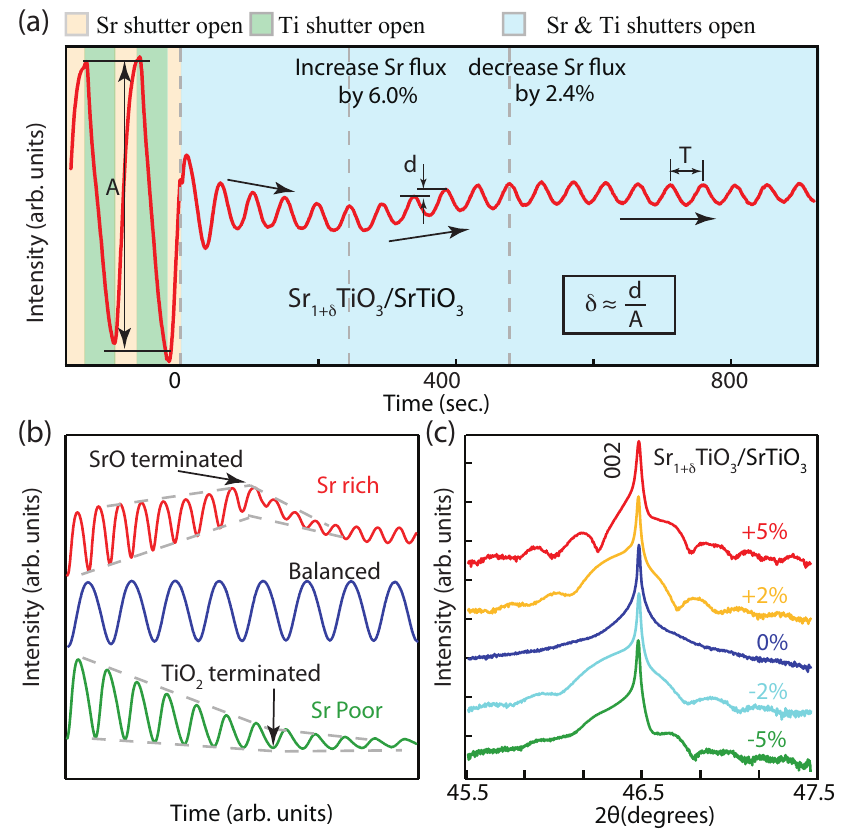}
\caption{(Color online) (a) RHEED oscillations during the co-deposition calibration process of homoepitaxial SrTiO$_3$ films. (b) Overall RHEED intensity drift for different Sr:Ti flux ratios. The kinks marked by arrows indicate the formation of full SrO (TiO$_2$) terminations due to the accumulation of excess Sr (Ti) during the growth. (c) XRD $\theta$-2$\theta$ scan patterns of homoepitaxial SrTiO$_3$  films grown by co-deposition method. 
\label{fig:co-dep}}
\end{center}
\end{figure}

Moreover, the rate of the overall intensity drift is found to be proportional to the accumulation rate of excess Sr or Ti atoms. As shown in Fig. 2b, the RHEED intensity reaches its maxima when a full SrO layer is formed on the surface due to the accumulation of excess Sr atoms. This is consistent with the fact that the SrO terminated surface gives the maximum RHEED intensity in the shuttered growth process. Accumulating more excess Sr atoms on the film surface results in a decrease of the overall diffraction intensity and the oscillation amplitude since the film surface crystalline quality becomes worse. In fact, this shares the same origin with the formation of "double peak" pattern in the shuttered growth process. Similarly, a minimum intensity is observed when a full TiO$_2$ terminated surface is formed by the accumulation of excess Ti atoms. 

Assuming a linear dependence of the overall intensity drift on the accumulation rate of excess Sr atoms, the amount of off-stoichiometry in Sr$_{1+\delta}$TiO$_3$ films can be quantified by Eq. (1), 
\begin{equation}
\delta\approx\frac{d}{A}
\end{equation} 
where, $d$ is the intensity drift during the growth of a unit cell (u.c.) and $A$ is the intensity difference between the RHEED diffraction of SrO-terminated and TiO$_2$-terminated SrTiO$_3$. For more reliable estimation, the $d$ values should be read after the oscillation amplitude and source temperature (flux) are both stabilised.

The validity of this first order estimation of the amount of off-stoichiometry $\delta$ using Eq. (1) was checked by comparing $\delta_\mathrm{RHEED}$ calculated using Eq(1) and $\delta_\mathrm{QCM}$ measured by a quartz crystal microbalance (QCM). From the data shown in Fig. 2, $\delta_\mathrm{RHEED}$ (2.4\%) agrees reasonably well with $\delta_\mathrm{QCM}$ (2.3\%). From our experience, Eq. (1) typically overestimates the amount of off-stoichiometry and $\delta_\mathrm{RHEED}$ can be up to 1.5$\delta_\mathrm{QCM}$ in some extreme cases, but this is still meaningful in practice. In contrast to the assumption of simple linear dependence of the diffraction intensity on the SrO coverage used in Eq. (1), more accurate estimation of $\delta$ should consider the exact sinusoidal-like dependence of the diffraction intensity on the SrO coverage.

Using co-deposition method, we can precisely control the film stoichiometry. As shown in Fig. 2c, a set of 40 nm thick epitaxial Sr$_{1+\delta}$TiO$_3$ films were grown on  (001) SrTiO$_3$ substrate using a co-deposition method. For the stoichiometric SrTiO$_3$ film, XRD $\theta$-2$\theta$ scans around the SrTiO3 (002) diffraction peak show no sign of peak splitting or Kiessig fringes, indicating the nearly perfect stoichiometry and crystalline structure of the homoepitaxial SrTiO$_3$ film~\cite{brooks2009growth}. In our experiments, the overall intensity drift of RHEED oscillations is found to be sensitive to Sr:Ti flux ratio. A minor change (0.1 $^{\circ}$C) of the Sr source temperature (about 0.3\% flux variation) can result in a clear drift of the overall oscillation intensity (not shown), indicating the Sr:Ti ratio can be controlled with a deviation less than 0.3\%. 


\begin{figure}
\begin{center}
\includegraphics{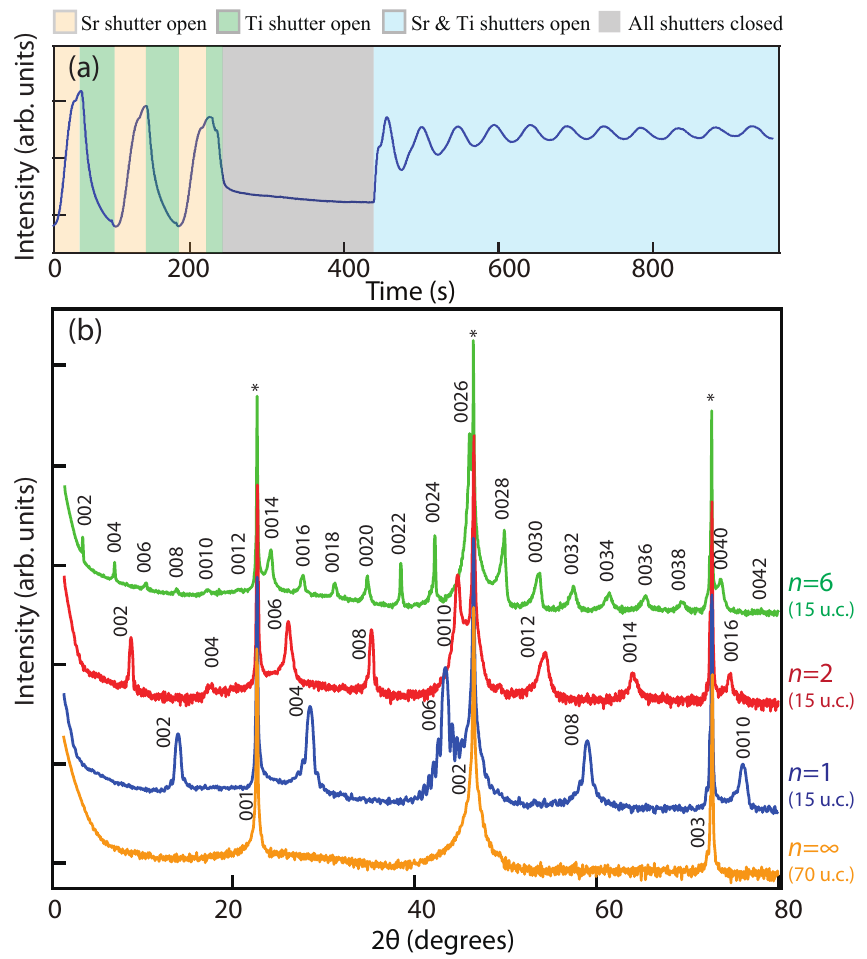}
\caption{(Color online) (a) RHEED oscillations for the calibration of homoepitaxial SrTiO$_3$ film. (b) XRD $\theta$-2$\theta$ scan patterns of Ruddlesden-Popper Sr$_{n+1}$Ti$_n$O$_{3n+1}$ films grown on (001) SrTiO$_3$ substrates by a combination of co-deposition calibration and shuttered final film growth. The asterisks denote the diffraction peaks from the substrate.  }
\label{fig:XRD_RP}
\end{center}
\end{figure}

One of the major advantages of the MBE technique is the extreme flexibility provided by the shuttered method, which is essential in growing heterostructures, superlattices and interfaces of exciting correlated quantum phases. To show that the  shutter times calibrated by the co-deposition technique can be directly used to grow complex structures using shuttered method, a series of RP strontium titanate (Sr$_{n+1}$Ti$_n$O$_{3n+1}$) films were grown on (001) SrTiO$_3$ substrates. In RP layered perovskites ($A_{n+1}B_nO_{3n+1}$), the $AB$O$_3$ blocks are interrupted by a rock-salt $A$O-$A$O double layers~\cite{ruddlesden1958compound}. The crystalline quality of RP structure is very sensitive to the dose of each sublayer~\cite{haeni2001epitaxial}. RP series host a variety of exciting properties including high temperature superconductors\cite{ANIEBACK:ANIE198914721}, colossal-magnetoresistance oxides\cite{Moritomo:1996ks},  spin-triplet superconductors\cite{Maeno:1994cm,Mackenzie.75.657}, and unconventional ferromagnets\cite{Callaghan1966}. 

As shown in Fig. 3,  XRD data of $n=1, 2, 6,$ RP films grown on (001) SrTiO$_3$ substrates exhibit clear and sharp diffraction peaks and clear Kiessig fringes, indicating the high quality of all RP films. The homoepitaxial SrTiO$_3$ (the $n=\infty$ member) film show no sign of peak splitting or  Kiessig fringes, indicating the close stoichiometric match of SrTiO$_3$ films. These results demonstrate that the co-deposition method and shuttered method can be combined to grow complex heterostructures and interfaces in an efficient and reliable manner. 

In summary, we show that the RHEED oscillation pattern in the co-deposition growth of SrTiO$_3$ is simple and clear by only exhibiting sensitive and monotonic dependance on the Sr:Ti flux ratio. By adjusting the Sr source temperature carefully, a stable RHEED oscillation pattern can be established and its period can be used as the precise shutter times for the shuttered growth of SrTiO$_3$ and SrTiO$_3$ based superlattices and interfaces. The combination of co-deposition calibration and shuttered final film growth is an efficient and reliable growth method for complex oxide superlattices, interfaces,  and will be helpful for the further improvement of other high-k oxide epitaxy layers as well.



{\bf Acknowledgements}

This work is supported by the the National Basic Research Programme of China (Grant No. 2015CB654901, 11574135, and 51672125). Y.F.N. is supported by the National Thousand-Young-Talents Program.


\end{document}